\documentclass[%
 reprint,
 amsmath,amssymb,
 aps,
 pra
]{revtex4-1}

\usepackage{graphicx}
\usepackage{dcolumn}
\usepackage{bm}

\begin{document}

\preprint{APS/123-QED}

\title{Decoherence in an exactly solvable qubit model with
 initial qubit-environment correlations}

\author{V.G.~Morozov}
\email{vladmorozov45@gmail.com}
 \affiliation{Moscow State Technical  University of Radioengineering,
  Electronics, and Automation,
 Vernadsky Prospect 78, 119454\ Moscow, Russia}

\author{S.~Mathey}
 \email{mathey@thphys.uni-heidelberg.de}
  \author{G.~R\"opke}
 \email{gerd.roepke@uni-rostock.de}
  \affiliation{
 University of Rostock, FB Physik,
   Universit\"atsplatz 3, D-18051\ Rostock, Germany
}

\date{\today}

\begin{abstract}
We study  a model of dephasing (decoherence) in a two-state quantum system
(qubit) coupled to a bath of harmonic oscillators. An exact analytic
solution for the reduced dynamics of a two-state system in this model has
been obtained previously for factorizing initial states of the combined
system. We show that
 the model admits exact solutions for a large class of correlated initial
 states which are typical in the  theory of quantum measurements.
 We derive exact expressions for the off-diagonal
 elements of the qubit density matrix,
  which hold for an arbitrary strength of coupling between the qubit and
  the bath. The influence of initial correlations on decoherence
 is considered for different bath spectral densities. Time behavior of the
 qubit entropy in the decoherence process is discussed.

\begin{description}
\item[PACS numbers] 03.65.Ta, 03.65.Yz
\end{description}
\end{abstract}

\maketitle

\section{\label{sec:Introd}INTRODUCTION}

There are two long-standing problems in the theory of open quantum systems
 --- memory effects and the influence of initial  statistical correlations
  on the system dynamics~\cite{BreuerPetr02,Weiss08}.
  These problems take on special significance in the theory of decoherence
 (i.e., the environment-induced destruction of quantum coherence),
  since  the decoherence time scale is usually much shorter than the
   time scales for other  relaxation processes in a system.
  The well-known  Nakajima-Zwanzig projection operator
  technique~\cite{Nakajima58,Zwanzig60} and its
 modifications~\cite{BreuerPetr02} provide formally exact non-Markovian
 master equations for almost arbitrary open systems and initial conditions.
 Unfortunately, it is impossible to solve these equations for more or
 less realistic models. Thanks to the work of many
 people (see, e.g.,
 Refs.~\cite{BreuerPetr02,VaccBreuer10,SmirneVacc10} and references therein),
  we possess today some partial advancements in constructing  reasonable
   approximations for  non-Markovian dynamics of open quantum systems, but a
systematic general approach is still lacking.

 The simplest systems involving many of fundamental features of quantum
  coherence are two-state systems. Such systems are important in their
  own right as the elementary carriers of
 quantum information (qubits)~\cite{Steane98,Bouwmeester00,Valiev05}.
In addition, some two-state models admit exact solutions.
  The latter fact is very useful because it allows one to gain valuable
   insight into general properties of the dynamics of decoherence and can
    serve as a step toward  consistent approximations for more
   complicated open systems. For instance,
   in Ref.~\cite{VaccBreuer10,SmirneVacc10} two exactly solvable
  models have been used to reconstruct the corresponding
  exact non-Markovian master equations and study in detail their
   memory kernels.
 It would be interesting to apply a similar approach to situations
  with non-negligible initial correlations between an open system and
   its environment.

   Recently~\cite{DajkaLucz10}, an exact analytic solution was obtained for a
    qubit model~\cite{Luczka90,Unruh95,Palma96} with initial
    qubit-environment correlations. The results demonstrate new
    interesting features of decoherence in the presence of initial
     correlations. Unfortunately, the initial state considered in
      Ref.~\cite{DajkaLucz10} seems to be somewhat artificial since it
      relies on the assumption that the qubit + environment system
      is initially prepared in a pure quantum state at zero temperature.
      In this paper we show that the model~\cite{Luczka90,Unruh95,Palma96}
 admits exact solutions for a large class of physically reasonable correlated
  initial states at finite temperatures and derive explicit expressions
  for the coherences (off-diagonal elements of the
  qubit density matrix).

  The distinctive feature of the dephasing
    model~\cite{Luczka90,Unruh95,Palma96}  is
that the average populations of the qubit states do not depend on time.
  In other words, there is no relaxation to complete equilibrium
   between the qubit and the environment, i.e.,
 the model is \textit{nonergodic\/}.
  Curiously enough this feature may be
  considered as an advantage of the model. Indeed, since
   the relaxation time $\tau^{}_{R}$ for dissipative processes
  is usually much larger than the decoherence (dephasing) time
   $\tau^{}_{D}$, the model~\cite{Luczka90,Unruh95,Palma96}
   corresponds to the limiting case
     $\tau^{}_{R}/\tau^{}_{D}\to\infty$, when
      all irrelevant corrections
 due to energy dissipation are removed.
  Thus one may expect that essential features of decoherence
  in this simple model will be similar to those in more
 involved but less tractable dissipative models.

The paper is structured as follows. In Sec.~II we give a brief
 description of the model and find exact solutions of equations
  of motion for all relevant operators. In Sec.~III
 we derive exact expressions for
    the elements of the qubit density
     matrix, which hold for correlated
     initial conditions. These expressions are used in
Sec.~IV to study possible regimes of decoherence for different
 bath spectral densities. The results are compared with
  the uncorrelated case.   Finally, in Sec. V  we derive
   an exact expression for the qubit entropy
  and discuss  its time behavior
       in the presence of initial qubit-environment correlations.

\section{THE MODEL}

We consider a simple version of a spin-boson model describing a two-state
system ($S$) coupled
 to a bath ($B$) of harmonic
  oscillators~\cite{Luczka90,Unruh95,Palma96,BreuerPetr02}.
In the ``spin'' representation for a qubit, the
 total Hamiltonian of the model is written as
(in our units $\hbar=1$)
 \begin{eqnarray}
    H&=& H^{}_{S}+ H^{}_{B} + H^{}_{\text{int}}\nonumber\\
   {}&=& \frac{\omega^{}_{0}}{2} \sigma^{}_{3}
    + \sum_{k}\omega^{}_{k}b^{\dagger}_{k}b^{}_{k}
   + \sigma^{}_{3}\sum_{k}
   \left(
        g^{}_{k} b^{\dagger}_{k} + g^{*}_{k}b^{}_{k}
   \right),
   \label{H-tot}
 \end{eqnarray}
where $\omega^{}_{0}$ is the energy difference between the excited state
 $|1\rangle$ and the ground state $|0\rangle$ of the qubit,
  and $\sigma^{}_{3}$ is one of the Pauli matrices
  $\sigma^{}_{1}, \sigma^{}_{2}, \sigma^{}_{3}$.
  Note that $\sigma^{}_{3}|1\rangle=|1\rangle$ and
$\sigma^{}_{3}|0\rangle=-|0\rangle$.
   Bosonic
  creation and annihilation
   operators $b^{\dagger}_{k}$ and
  $b^{}_{k}$ correspond to the
 $k$th bath mode with frequency $\omega^{}_{k}$.

 Suppose that at time $ t=0$ the state of the total system is described
  by some initial density matrix
  $\varrho(0)$. Then at time $t$ the density matrix is given by
   \[
  \varrho(t)=\exp\left(-iHt\right)\varrho(0)\exp\left(iHt\right).
   \]
 Our main interest is with the
 reduced density matrix of the qubit
   \begin{equation}
   \varrho^{}_{S}(t)=\text{Tr}^{}_{B}\varrho(t),
   \label{rhoS-def}
 \end{equation}
where $\text{Tr}^{}_{B}$ denotes the trace taken over the bath degrees of
freedom. It is easy to see that
  \begin{equation}
   \begin{array}{l}
   \langle 0|\varrho^{}_{S}(t)|0 \rangle=
   \frac{1}{2}\left\{ 1- \langle \sigma^{}_{3}(t)\rangle\right\},
    \\[5pt]
   \langle 1|\varrho^{}_{S}(t)|1 \rangle=
   \frac{1}{2}\left\{ 1+ \langle \sigma^{}_{3}(t)\rangle\right\},
   \\[5pt]
  \langle 0|\varrho^{}_{S}(t)|1 \rangle= \langle\sigma^{}_{+}(t)\rangle,
   \quad
   \langle 1|\varrho^{}_{S}(t)|0 \rangle=\langle\sigma^{}_{-}(t)\rangle,
    \end{array}
    \label{DMatS-sig}
  \end{equation}
where $\sigma^{}_{\pm}=\left(\sigma^{}_{1}\pm i\sigma^{}_{2}\right)/2$.
Here and in the following the symbol $\langle A(t)\rangle$ stands for the
average value of a Heisenberg picture operator calculated with the initial
density matrix of the total system:
  \[
 \langle A(t)\rangle= {\rm Tr}\left\{
 \exp\left(iHt\right)A\exp\left(-iHt\right)\varrho(0)
  \right\}.
  \]
The notation $\langle A\rangle$ will be used for averages at $t=0$.

In the model (\ref{H-tot}), equations of motion for all relevant
 operators can be solved exactly. Technical details are summarized in
Appendix~\ref{App-A-Heis}. Here we quote the results.
 The time-dependent bath operators are given by
 \begin{equation}
  \begin{array}{l}
  \displaystyle
  b^{}_{k}(t)={\rm e}^{-i\omega^{}_{k}t}\left[
   b^{}_{k}+ \frac{\sigma^{}_{3}}{2}\, \alpha^{}_{k}(t)
    \right],
  \\[10pt]
  \displaystyle
 b^{\dagger}_{k}(t)={\rm e}^{i\omega^{}_{k}t}\left[
   b^{\dagger}_{k}+  \frac{\sigma^{}_{3}}{2}\,\alpha^{*}_{k}(t)
   \right],
   \end{array}
    \label{b(t)}
 \end{equation}
with
  \begin{equation}
   \alpha^{}_{k}(t)=
   2g^{}_{k}\frac{1-{\rm e}^{i\omega^{}_{k}t}}{\omega^{}_{k}}\, ,
   \label{alpha}
  \end{equation}
 and the qubit operators $\sigma^{}_{\pm}(t)$ can be written as
  \begin{equation}
    \sigma^{}_{\pm}(t)=
    \exp\left[   \pm i\omega^{}_{0}t \mp R(t)
        \right] \sigma^{}_{\pm}\, ,
   \label{sig-pm(t)}
  \end{equation}
where the operator $R(t)$  acts only on the bath states:
   \begin{equation}
 R(t)=
     \sum_{k}\left[
     \alpha^{}_{k}(t) b^{\dagger}_{k}-\alpha^{*}_{k}(t) b^{}_{k}
             \right].
    \label{R(t)}
   \end{equation}
Since  $\sigma^{}_{3}$ commutes with the Hamiltonian (\ref{H-tot}), we
have $\sigma^{}_{3}(t)=\sigma^{}_{3}$, so that the
  populations $\langle 0|\varrho^{}_{S}(t)|0 \rangle$
 and $\langle 1|\varrho^{}_{S}(t)|1 \rangle$ do not depend on time.

\section{EXACT SOLUTIONS FOR THE COHERENCES\label{SecSolution}}

\subsection{Uncorrelated initial state}

As an introduction to our subsequent development, we first briefly review
the calculation of
 the coherences $\langle\sigma^{}_{\pm}(t)\rangle$ in the case that the
  qubit and the
bath are initially uncorrelated, and the bath is in thermal equilibrium at
some temperature $T$~\cite{BreuerPetr02,Luczka90,Unruh95,Palma96}. In this
case the initial density matrix of the total system is a direct product
   \begin{equation}
    \varrho(0)=\varrho^{}_{S}(0)\otimes \varrho^{}_{B},
     \qquad
     \left. \varrho^{}_{B}=\text{e}^{-\beta H^{}_{B}}\right/Z^{}_{B}\, ,
     \label{Init-fact}
   \end{equation}
where $\beta=1/k^{}_{\text{B}}T$, and $Z^{}_{B}$ is the bath partition
function. Note that $\varrho^{}_{S}(0)$ may be a pure state as well as a
mixed state of the qubit.

Using  expressions (\ref{sig-pm(t)}) and (\ref{Init-fact}), one obtains
   \begin{equation}
    \langle \sigma^{}_{\pm}(t)\rangle =
     \langle{\sigma^{}_{\pm}}\rangle\,\text{e}^{\pm i\omega^{}_{0}t}\,
      \text{e}^{-\gamma(t)}
    \label{sig-factor}
   \end{equation}
 with the \textit{decoherence function\/} $\gamma(t)$ defined as
  \begin{eqnarray}
  \gamma(t)&=& -
   \ln\,\left\langle
    \text{e}^{\mp R(t)}\right\rangle_{B}
   \nonumber\\
    {}&=&
    - \sum_{k}\ln \left\langle
       \exp\left[
       \alpha^{}_{k}(t) b^{\dagger}_{k} -
           \alpha^{*}_{k}(t) b^{}_{k}
           \right]\right\rangle^{}_{B},
   \label{Gamma(t)}
  \end{eqnarray}
 where the symbol $\langle \ldots \rangle^{}_{B}$ denotes averages taken
  with the bath
distribution $\varrho^{}_{B}$. After straightforward algebra (see, e.g.,
Ref.~\cite{BreuerPetr02}) one finds
   \begin{equation}
 \gamma(t)=  \int^{\infty}_{0} d\omega\,
  J(\omega)\,\coth\left(\beta\omega/2\right)\,
  \frac{1-\cos\omega t}{\omega^{2}}\, ,
      \label{Gamma(t)-contin}
   \end{equation}
 were the continuum limit of the bath modes is performed, and the spectral
 density $J(\omega)$ is introduced by the rule
   \begin{equation}
    \sum_{k} 4|g^{}_{k}|^{2}\,f(\omega^{}_{k})=
     \int^{\infty}_{0} d\omega\, J(\omega)f(\omega).
     \label{J(omega)}
   \end{equation}
Expression (\ref{Gamma(t)-contin}) is the exact result for the
 decoherence function in the
model (\ref{H-tot}) under the uncorrelated  initial condition
(\ref{Init-fact}).

\subsection{Correlated initial states}

We turn now to the coherences
 $\langle \sigma^{}_{\pm}(t)\rangle$ for correlated initial states.
 First of all, we have to specify the initial density matrix
  $\varrho(0)$. Generally speaking, one may imagine
    a variety of different forms of this density matrix, but we will
 restrict our consideration to initial states
   \begin{equation}
   \varrho(0)=
    \frac{1}{Z}\,
    \sum_{m}\Omega^{}_{m}\,\text{e}^{-\beta H}\Omega^{\dagger}_{m},
     \label{Init-corr}
   \end{equation}
where operators $\Omega^{}_{m}$ act on the qubit states, and the partition
function $Z$ ensures the normalization of $\varrho(0)$. Such density
matrices are used, for instance, to describe the preparation of a system
by means of a quantum measurement~\cite{BreuerPetr02}. They also arise in
a natural way in the calculation of
 correlation functions of open quantum  systems. In the
 theory of decoherence, of special interest are initial density
 matrices~\cite{BreuerPetr02,Weiss08}
    \begin{equation}
    \varrho(0)=
    \frac{1}{Z}\,
    P^{}_{\psi}\text{e}^{-\beta H}P^{}_{\psi},
     \label{Pure-corr}
    \end{equation}
where $P^{}_{\psi}=|\psi\rangle \langle\psi|$ is the projector onto a pure
quantum state $|\psi\rangle$. In the usual representation, we have
  \begin{equation}
 |\psi\rangle= a^{}_{0}|0\rangle + a^{}_{1}|1\rangle
    \label{Pure-psi}
  \end{equation}
with $|a^{}_{0}|^{2} + |a^{}_{1}|^{2}=1$.
  The projector $P^{}_{\psi}$ can also be written as
          \begin{equation}
  P^{}_{\psi}=\frac{1}{2}\left(1+ \vec{\sigma}\cdot\vec{p}\,\right),
   \qquad
    |\vec{p}\,|=1,
    \label{P-sigma}
   \end{equation}
 where the components of
 $\vec{p}$ are easily expressed in terms of the
amplitudes $a^{}_{0}$ and $a^{}_{1}$.
 The density matrix (\ref{Pure-corr}) corresponds to the so-called
 selective quantum measurement~\cite{BreuerPetr02} and is
especially suited to study the environment-induced destruction of quantum
coherence.

In contrast to density matrices of the form (\ref{Init-fact}), the density
matrices (\ref{Init-corr}) and (\ref{Pure-corr}) contain
  the \textit{total\/} Hamiltonian of the  system and, consequently,
  describe initial qubit-bath correlations.
 If we neglect the interaction
 term $H^{}_{\text{int}}$, replacing
 $H$ by $H^{}_{0}=H^{}_{S} + H^{}_{B}$, then we immediately
recover an uncorrelated state (\ref{Init-fact}). It is interesting to note
that the density matrix (\ref{Pure-corr}) can be written as
  \begin{equation}
   \varrho(0)=P^{}_{\psi}\otimes \varrho^{}_{B}(|\psi\rangle),
   \label{Init-corr-prod}
  \end{equation}
where $\varrho^{}_{B}(|\psi\rangle)$ plays the role of the initial density
matrix of the bath and, at the same time, is a functional of
 $|\psi\rangle$:
   \begin{equation}
 \varrho^{}_{B}(|\psi\rangle)=
  \frac{\langle\psi|\exp(-\beta H)|\psi\rangle}
       {\text{Tr}^{}_{B}\langle\psi|\exp(-\beta H)|\psi\rangle}.
     \label{Bath-psi}
   \end{equation}
In the model under consideration, this functional can be obtained
explicitly. First we note that the Hamiltonian (\ref{H-tot})
 satisfies
      \begin{equation}
   \begin{array}{l}
   \text{e}^{-\beta H}|0\rangle=
   \text{e}^{\beta\omega^{}_{0}/2}\,
   \text{e}^{-\beta H^{(-)}_{B}}\otimes |0\rangle,
   \\[3pt]
   \text{e}^{-\beta H}|1\rangle=
   \text{e}^{-\beta\omega^{}_{0}/2}\,
   \text{e}^{-\beta H^{(+)}_{B}} \otimes |1\rangle\, ,
   \end{array}
     \label{H-prop}
  \end{equation}
 where
   \begin{equation}
   H^{(\pm)}_{B}=
   \sum_{k} \omega^{}_{k} b^{\dagger}_{k} b^{}_{k}
  \pm  \sum_{k}
   \left(g^{}_{k} b^{\dagger}_{k}+g^{*}_{k} b^{}_{k}\right).
    \label{H-pm}
   \end{equation}
Then, with Eqs.~(\ref{H-prop}), the initial bath density matrix
(\ref{Bath-psi}) is manipulated to
  \begin{equation}
 \varrho^{}_{B}(|\psi\rangle)=
    \frac{|a^{}_{0}|^{2}\text{e}^{\beta\omega^{}_{0}/2}
           \text{e}^{-\beta H^{(-)}_{B}} +
          |a^{}_{1}|^{2} \text{e}^{-\beta\omega^{}_{0}/2}
           \text{e}^{-\beta H^{(+)}_{B}}}
           {
            |a^{}_{0}|^{2}\text{e}^{\beta\omega^{}_{0}/2} Z^{(-)}_{B} +
            |a^{}_{1}|^{2} \text{e}^{-\beta\omega^{}_{0}/2} Z^{(+)}_{B}
           }\, ,
     \label{Bath-psi-expl}
  \end{equation}
where
 \begin{equation}
 Z^{(\pm)}_{B}=\text{Tr}^{}_{B}
  \exp\left[
       -\beta H^{(\pm)}_{B}
       \right].
       \label{Z-pm-B}
 \end{equation}
Thus, although the density matrices (\ref{Init-fact}) and
(\ref{Init-corr-prod}) are similar in form, the latter involves initial
qubit-bath correlations. As shown below, these
 correlations can significantly affect the dynamics of the total system.

\subsection{Time evolution of correlated initial states}

The structure of the coherences $\langle \sigma^{}_{\pm}(t)\rangle$ can be
 investigated for arbitrary operators $\Omega^{}_{m}$
 in Eq.~(\ref{Init-corr}). First we write
   \begin{eqnarray*}
 \langle \sigma^{}_{\pm}(t)\rangle &=&
  \frac{1}{Z}\sum_{m}
   \text{Tr}\left[
        \Omega^{\dagger}_{m}\sigma^{}_{\pm}(t)\Omega^{}_{m}
         \text{e}^{-\beta H}
            \right]
            \\
            &=&
 \frac{\text{e}^{\pm i\omega^{}_{0}t}}{Z}\sum_{m}
   \text{Tr}^{}_{B}\!\left\{
   \text{Tr}^{}_{S}\left[
        \Omega^{\dagger}_{m}\sigma^{}_{\pm}\Omega^{}_{m}
          \text{e}^{\mp R(t)}\text{e}^{-\beta H}
            \right]
            \right\}.
   \end{eqnarray*}
Using Eqs.~(\ref{H-prop}), the above expression
 is recast into
  \begin{widetext}
  \begin{eqnarray}
   & &
   \hspace*{-50pt}
   \langle\sigma^{}_{\pm}(t)\rangle=
   \frac{ \text{e}^{\pm i\omega^{}_{0}t}}{Z}\,
  \sum_{m}\left\{ \text{e}^{\beta\omega^{}_{0}/2}\,
   \langle 0|\Omega^{\dagger}_{m}\sigma^{}_{\pm}\Omega^{}_{m} |0\rangle
   \, \text{Tr}^{}_{B}
   \left( \text{e}^{\mp R(t)} \text{e}^{-\beta H^{(-)}_{B}} \right)
   \right.
   \nonumber\\
   & &
   \hspace*{90pt}
   \left.
   {}+
 \text{e}^{-\beta\omega^{}_{0}/2}\,
   \langle 1|\Omega^{\dagger}_{m}\sigma^{}_{\pm}\Omega^{}_{m} |1\rangle
   \, \text{Tr}^{}_{B}
   \left( \text{e}^{\mp R(t)} \text{e}^{-\beta H^{(+)}_{B}} \right)
   \right\}
  \label{sig-Omega}
\end{eqnarray}
  \end{widetext}
with the partition function
  \begin{eqnarray}
   Z &=&
    \sum_{m}\left\{
    \text{e}^{\beta\omega^{}_{0}/2}\,
   \langle 0|\Omega^{\dagger}_{m}\Omega^{}_{m} |0\rangle
   \,  Z^{(-)}_{B}
            \right.
    \nonumber\\
     & &
            \left.
   {}+
    \text{e}^{-\beta\omega^{}_{0}/2}\,
   \langle 1|\Omega^{\dagger}_{m}\Omega^{}_{m} |1\rangle
   \, Z^{(+)}_{B}
             \right\}.
   \label{Z-Omega}
  \end{eqnarray}
We now note that the traces $\text{Tr}^{}_{B}\,(\ldots)$ in
 Eq. (\ref{sig-Omega}) can be simplified considerably by
  a unitary transformation of  $H^{(\pm)}_{B}$ and $R(t)$
   with
    \[
    U^{}_{\pm}=\exp\left\{
    \pm  \sum_{k}\left(
    \frac{g^{}_{k}}{\omega^{}_{k}}\,b^{\dagger}_{k}
    -\frac{g^{*}_{k}}{\omega^{}_{k}}\,b^{}_{k}
    \right)
   \right\}.
    \]
It is easy to verify that
  \begin{equation}
    \begin{array}{l}
   \displaystyle
       U^{}_{\pm}H^{(\pm)}_{B}U^{-1}_{\pm}= H^{}_{B}
    -  \sum_{k} \frac{|g^{}_{k}|^{2}}{\omega^{}_{k}},
    \\[5pt]
   U^{}_{\pm}R(t)U^{-1}_{\pm}=
    R(t) \pm\,i\Phi(t),
   \end{array}
    \label{Trans-HR}
  \end{equation}
where we have introduced the $c$-number function
 \begin{equation}
 \Phi(t)=
 \sum_{k} \frac{4|g^{}_{k}|^{2}}{\omega^{2}_{k}}\,
    \sin(\omega^{}_{k}t)=
  \int^{\infty}_{0} d\omega\,
  J(\omega)\,\frac{\sin \omega t}{\omega^{2}}
  \label{Phi-def}
 \end{equation}
with the same spectral density $J(\omega)$ as in
 Eq.~(\ref{Gamma(t)-contin}). Using the transformation properties
  (\ref{Trans-HR}) in Eqs.~(\ref{sig-Omega}) and (\ref{Z-Omega}),
  we obtain the final expression for the coherences,
   which can be conveniently written as
     \begin{widetext}
    \begin{equation}
   \langle\sigma^{}_{\pm}(t)\rangle=\langle \sigma^{}_{\pm}\rangle\,
   \text{e}^{\pm i\omega^{}_{0}t}\,\text{e}^{-\gamma(t)}
 \frac{
  \sum_{m}\left\{
   \langle 0|\Omega^{\dagger}_{m}\sigma^{}_{\pm}\Omega^{}_{m}|0\rangle
    \text{e}^{\beta\omega^{}_{0}/2}\, \text{e}^{\pm i\Phi(t)}
   +
   \langle 1|\Omega^{\dagger}_{m}\sigma^{}_{\pm}\Omega^{}_{m}|1\rangle
  \text{e}^{-\beta\omega^{}_{0}/2}\, \text{e}^{\mp i\Phi(t)}
  \right\}
    }
    {
    \sum_{m}\left\{
   \langle 0|\Omega^{\dagger}_{m}\sigma^{}_{\pm}\Omega^{}_{m}|0\rangle
 \text{e}^{\beta\omega^{}_{0}/2}
   +
   \langle 1|\Omega^{\dagger}_{m}\sigma^{}_{\pm}\Omega^{}_{m}|1\rangle
  \text{e}^{-\beta\omega^{}_{0}/2}
  \right\}
   },
   \label{sig-fin-gen}
 \end{equation}
 \end{widetext}
where $\langle \sigma^{}_{\pm}\rangle$ are initial values of the
coherences.

Formulas (\ref{Phi-def}) and (\ref{sig-fin-gen}) give an exact result for
the decoherences in the model (\ref{H-tot}) with initial correlated states
of the form (\ref{Init-corr}). Expression (\ref{sig-fin-gen}) is
considerably simplified in the special case when  the initial density
matrix is given by Eq.~(\ref{Pure-corr}). After some algebra which we
omit, we obtain
  \begin{eqnarray}
   & &
 \hspace*{-15pt}
     \langle\sigma^{}_{\pm}(t)\rangle =
   \langle \sigma^{}_{\pm}\rangle\, \text{e}^{\pm i\omega^{}_{0}t}
  \text{e}^{-\gamma(t)}\,
 \bigg\{ \cos[\Phi(t)]
  \nonumber \\
   & &
   {} \pm i\,
  \frac{\sinh(\beta\omega^{}_{0}/2) -\langle\sigma^{}_{3}\rangle
   \cosh(\beta\omega^{}_{0}/2)
     }
       { \cosh(\beta\omega^{}_{0}/2)-\langle\sigma^{}_{3}\rangle
   \sinh(\beta\omega^{}_{0}/2)
   }
  \,\sin[\Phi(t)]
   \bigg\},
   \label{sig-P}
 \end{eqnarray}
where $\langle \sigma^{}_{3}\rangle=|a^{}_{1}|^{2}- |a^{}_{0}|^{2}$. This
expression can be represented in a more transparent form
  \begin{equation}
       \langle\sigma^{}_{\pm}(t)\rangle =
   \langle \sigma^{}_{\pm}\rangle\, \text{e}^{\pm i[\omega^{}_{0}t+\chi(t)]}
 \text{e}^{-\widetilde\gamma(t)},
      \label{sig-tilde}
  \end{equation}
where
  \begin{equation}
\widetilde\gamma(t)=\gamma(t) + \gamma^{}_{\text{corr}}(t)
  \label{Gamma-tilde}
   \end{equation}
is the modified decoherence function which includes the correlation
contribution
 \begin{eqnarray}
  & &
  \hspace*{-16pt}
 \gamma^{}_{\text{corr}}(t)
  \nonumber\\
   & &
 \hspace*{-16pt}
  {}=
   -\frac{1}{2}\,\ln
   \left[
    1 -\frac{\left(1- \langle\sigma^{}_{3}\rangle^{2}\right)
        \sin^{2}[\Phi(t)]
            }
            {\left[\cosh(\beta\omega^{}_{0}/2)-\langle\sigma^{}_{3}\rangle
             \sinh(\beta\omega^{}_{0}/2)\right]^{2}
            }
   \right]\! ,
    \label{Gamma-corr}
 \end{eqnarray}
and $\chi(t)$ is the time-dependent phase shift with
  \begin{equation}
  \tan[\chi(t)]=
  \frac{\sinh(\beta\omega^{}_{0}/2) -\langle\sigma^{}_{3}\rangle
   \cosh(\beta\omega^{}_{0}/2)
     }
       { \cosh(\beta\omega^{}_{0}/2)-\langle\sigma^{}_{3}\rangle
   \sinh(\beta\omega^{}_{0}/2)
   }
  \,\tan[\Phi(t)].
   \label{chi(t)}
  \end{equation}
It is a straightforward matter to derive a formula analogous to
Eq.~(\ref{sig-tilde}) for
 more general initial states (\ref{Init-corr}) because each of the operators
 $\Omega^{}_{m}$ in Eq.~(\ref{sig-fin-gen}) can always be
represented as $\Omega^{}_{m}=c^{}_{m} + \vec{\sigma}\cdot\vec{p}^{}_{m}$
with some (in general, complex) $c^{}_{m}$ and $\vec{p}^{}_{m}$. The
resulting  expressions for $\gamma^{}_{\text{corr}}(t)$ and $\chi(t)$ are
rather cumbersome and will not be given here. From now on we restrict our
discussion to  Eqs.~(\ref{sig-tilde})\,--\,(\ref{chi(t)}).

 \section{Regimes of decoherence for different bath spectral densities}

 As already noted, the model (\ref{H-tot}) is nonergodic,
 i.e., it does not describe the establishment of complete thermal
 equilibrium between the qubit and the bath since
 $\langle\sigma^{}_{3}(t)\rangle=\text{const}$.
In this connection it is of interest to
  investigate the long-time behavior of the decoherence function
  (\ref{Gamma-tilde}). Suppose, for instance, that $\widetilde{\gamma}(t)$
 is a bounded function. Then, in the limit as $t\to \infty$,
 the averages $\langle\sigma^{}_{\pm}(t)\rangle$
 do not tend to zero, so that the destruction of quantum coherence
 is ``incomplete''. Physically, in this case the final stage of
 decoherence is determined by slow processes involving exchange
 of energy between the qubit and the environment, which are not
 included into the Hamiltonian (\ref{H-tot}).  On the other hand, if
 $\widetilde{\gamma}(t)\to\infty$ as $t\to\infty$, then
  $\langle\sigma^{}_{\pm}(t)\rangle\to 0$ showing complete decoherence
  within the framework of the model under consideration.

Note that the correlation corrections in Eq.~(\ref{Gamma-corr}) [or, for a
 more general case, in Eq.~(\ref{sig-fin-gen})] are always bounded.
 This means that initial correlations alone cannot lead to complete
 decoherence. Let us now turn to the ``dynamical'' part of the
 decoherence function given by
Eq.~(\ref{Gamma(t)-contin}). It can be written conveniently as a sum
  \begin{equation}
  \gamma(t)=  \gamma^{}_{\text{vac}}(t)+ \gamma^{}_{\text{th}}(t),
    \label{Gamma-sum}
  \end{equation}
where
       \begin{equation}
   \gamma^{}_{\text{vac}}(t)=  \int^{\infty}_{0} d\omega\,
  J(\omega)\,
  \frac{1-\cos\omega t}{\omega^{2}}
   \label{Gamma(t)-vac}
   \end{equation}
is the contribution to the decoherence function from vacuum fluctuations
in the bath, and
         \begin{eqnarray}
   \gamma^{}_{\text{th}}(t) &=&  \int^{\infty}_{0} d\omega\,
  J(\omega)
  \left[
 \coth\left(\beta\omega/2\right)-1
  \right]
  \frac{1-\cos\omega t}{\omega^{2}}
 \nonumber\\
  {}&=&
 2\int^{\infty}_{0} d\omega\,\frac{J(\omega)}{{\rm e}^{\beta\omega}-1}\,
  \frac{1-\cos\omega t}{\omega^{2}}
   \label{Gamma(t)-th}
   \end{eqnarray}
is the temperature dependent thermal contribution. We have obvious
inequalities
  \begin{subequations}
     \label{Gamma-suff}
    \begin{eqnarray}
    & &
   \displaystyle
 \gamma^{}_{\text{vac}}(t)\leq
    2\int^{\infty}_{0} d\omega\,
  \frac{ J(\omega)}{\omega^{2}}\, ,
      \label{Gamma-suff-vac}
  \\[5pt]
  & &
   \displaystyle
 \gamma^{}_{\text{th}}(t)\leq
  4\int^{\infty}_{0} d\omega\,
  \frac{J(\omega)}{\omega^{2}\left({\rm e}^{\beta\omega}-1\right)}\, ,
 \label{Gamma-suff-th}
  \end{eqnarray}
   \end{subequations}
which show that  $\gamma^{}_{\text{vac}}(t)$ and
 $\gamma^{}_{\text{th}}(t)$ are bounded if the integrals converge.

Another important question is whether, and under what conditions, the
 total decoherence function $\widetilde{\gamma}(t)$ tends to
 a definite limit as $t\to \infty$.
  Necessary conditions for different terms in Eq.~(\ref{Gamma-tilde}) to
have definite long-time limits can be derived by using
  relation~\cite{Doetsch74}
   \begin{equation}
 \lim_{t\to\infty} f(t)= \lim_{\varepsilon\to +0}
 \varepsilon \int^{\infty}_{0} dt\,{\rm e}^{-\varepsilon t} f(t),
   \label{Abel}
 \end{equation}
which is valid if the limit on the left-hand side exists. When applied to
 functions (\ref{Gamma(t)-vac}),
 (\ref{Gamma(t)-th}), and (\ref{Phi-def}), Eq.~(\ref{Abel}) gives
    \begin{subequations}
     \label{Gamma-infty}
    \begin{eqnarray}
    & &
    \lim_{t\to\infty}\gamma^{}_{\text{vac}}(t)=
\int^{\infty}_{0} d\omega\,
  \frac{J(\omega)}{\omega^{2}}\, ,
 \label{Gamma-infty-vac}\\[5pt]
   & &
  \lim_{t\to\infty}\gamma^{}_{\text{th}}(t)=
  2\int^{\infty}_{0}
   d\omega\,
   \frac{J(\omega)}{\omega^{2}\left({\rm e}^{\beta\omega}-1\right)}\, ,
  \label{Gamma-infty-th}\\[5pt]
   & &
    \lim_{t\to\infty} \Phi(t)=
    \frac{\pi}{2}\,\lim_{\omega\to 0}\frac{J(\omega)}{\omega}\, .
\label{Phi-infty}
   \end{eqnarray}
  \end{subequations}

 To go beyond relations (\ref{Gamma-suff}) and (\ref{Gamma-infty}),
  one needs some information
 about the bath spectral density. Generally speaking, its form can be
 obtained from a fully macroscopic analysis of the system-bath
  interactions leading to the spin-boson
 model (\ref{H-tot}). It would be beyond the scope of this paper to
 present a detailed discussion of such derivations.
 In many cases of practical interest
  (see, e.g., Refs.~\cite{Palma96,Legett87}),
 $J(\omega)$ may be considered to be
 a reasonably smooth function
 which has a power-law behavior  $J(\omega)\propto \omega^{s}$\ ($s>0$)
  at frequencies much less than some ``cutoff'' frequency
   $\Omega$, characteristic of the bath modes. In the limit
    $\omega\to\infty$, $J(\omega)$ is assumed to fall off at least
    as some negative power of $\omega$.
    Then, on dimensional grounds, we may write
       \begin{equation}
   J(\omega)= \lambda^{}_{s}\Omega^{1-s}\,\omega^{s}\,
    F(\omega/\Omega),
    \label{J-model}
  \end{equation}
where $\lambda^{}_{s}$ is a dimensionless coupling constant, and a cutoff
function satisfies
   \begin{equation}
   F(0)=1, \qquad
    \lim^{}_{\omega/\Omega\to\infty} F(\omega/\Omega)=0.
     \label{F-def}
   \end{equation}
  The case $s=1$ is usually called the ``ohmic'' case, the  case $s>1$
  ``superohmic'', and the case $0<s<1$ ``subohmic''.

As is easy to see, the convergence of the integrals in
Eqs.~(\ref{Gamma-suff})  and (\ref{Gamma-infty})
 depends crucially on the low-frequency behavior of $J(\omega)$.
  Assuming a power law  for  $J(\omega)$ in this region,
 we can make some conclusions about
 possible regimes of decoherence in the model (\ref{H-tot}).
  Let us first of all note that both integrals in
  inequalities~(\ref{Gamma-suff}) converge for $s>2$.
   Therefore, in this case the
  total decoherence function $\widetilde{\gamma}(t)$ is bounded, so that
   the above mentioned regime of ``incomplete decoherence'' takes place.
 It is also seen from Eqs.~(\ref{Gamma-suff}) that, in the superohmic
 case with $1<s\leq 2$, the thermal term  $\gamma^{}_{\text{th}}(t)$ is the only
 contribution to the total decoherence function which may
  diverge as $t\to \infty$. Finally, Eq.~(\ref{Phi-infty}) shows that
  the correlation term $\gamma^{}_{\text{corr}}(t)$
   may have a nonzero long-time limit only
  in the ohmic case ($s=1$).

The virtue of the above conclusions is that they apply to all spectral
densities of the form (\ref{J-model}), but  they say nothing about the
regime of decoherence for $0<s \leq 2$. It is reasonable to expect that
 in this range of values of the parameter $s$ we are dealing with the
 regime of ``complete decoherence'', since the vacuum term
  $\gamma^{}_{\text{vac}}(t)$ may diverge as
  $t\to\infty$ for $0<s\leq 1$ and  the thermal
  term $\gamma^{}_{\text{th}}(t)$ for $0<s\leq 2$.
   We could not determine, however, a general sufficient
  condition for the cutoff function in Eq.~(\ref{J-model})
  to ensure that $\gamma^{}_{\text{vac}}(t)$ and
  $\gamma^{}_{\text{th}}(t)$ behave in this manner.
  Note, however, that this question is largely academic since
 the model specified by the Hamiltonian (\ref{H-tot}), on its own,
  is of most physical interest
 in the low-temperature range ($\Omega\beta\gg 1$) where dissipative
 processes are irrelevant and  the bath dynamics is completely determined by
  low-frequency modes with $\omega^{}_{k}\ll \Omega$.
 For this temperature range, we show in Appendix~\ref{App-C-Der} that,
 under assumptions (\ref{J-model}) and (\ref{F-def}),  the thermal term
 $\gamma^{}_{\text{th}}(t)$ is a steadily increasing function
 of time for $0<s\leq 2$, and hence  $\gamma^{}_{\text{th}}(t)\to \infty$
 as $t\to \infty$. This means that the regime of ``complete
 decoherence'' takes place for all $0<s\leq 2$.

Up to this point we have been concerned with
 those features of decoherence which do not depend on
 the form of the bath spectral density or, in particular,
 on the form of the cutoff function in Eq.~(\ref{J-model}).
 For specific cases, functions (\ref{Gamma-corr}), (\ref{Gamma(t)-vac}),
 and (\ref{Gamma(t)-th}) can be evaluated numerically.
  However, to gain more insight into the overall picture of decoherence, it will be
 instructive to consider a particular choice of the cutoff function in
 Eq.~(\ref{J-model}) for which all the quantities of interest can be evaluated
  exactly. We shall take for the bath spectral density the expression
    \begin{equation}
   J(\omega)= \lambda^{}_{s}\Omega^{1-s}\,\omega^{s}\,
    {\rm e}^{-\omega/\Omega}\, ,
    \label{J-model-exp}
  \end{equation}
which is most commonly used in the theory of spin-boson
 systems~\cite{BreuerPetr02,Luczka90,Unruh95,Palma96,Legett87}.

\begin{figure}[htb] 
\begin{center}
{\includegraphics[width=7.5cm]{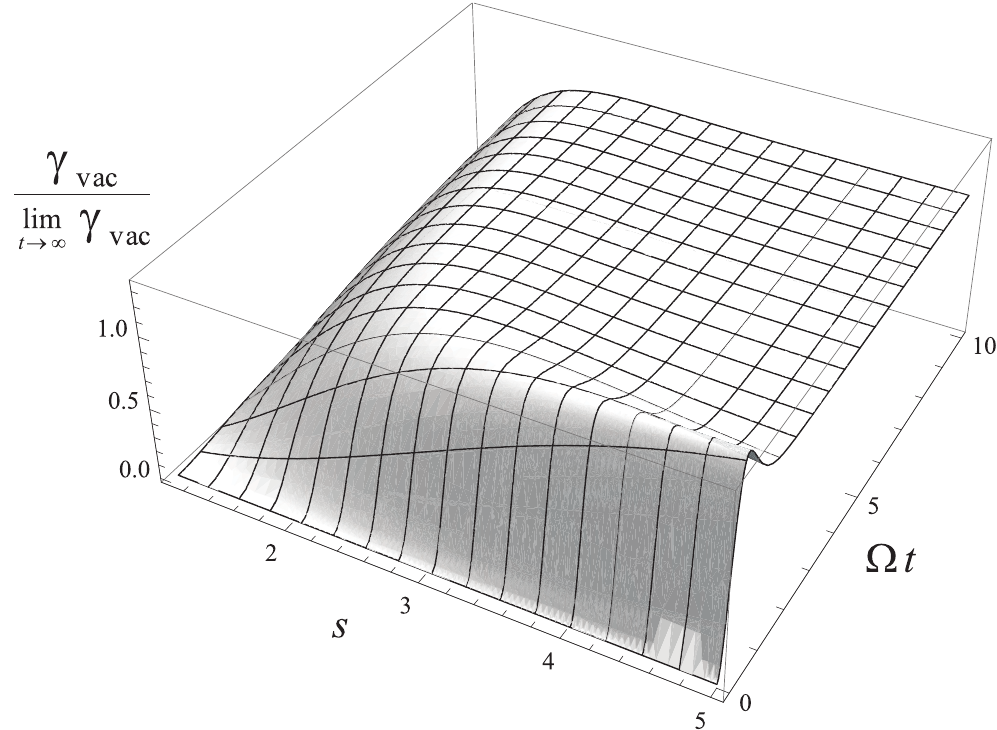}}
 \end{center}
 \caption{Time dependence of the vacuum contribution to the decoherence
  function in the superohmic case ($s>1$).}
  \label{Fig-1-r}
\end{figure}

Let us start with the vacuum term (\ref{Gamma(t)-vac}) in the decoherence
function. Substituting here the spectral density from
Eq.~(\ref{J-model-exp}) and doing standard integrals, we get
 \begin{subequations}
  \label{Gamma-vac-s}
 \begin{eqnarray}
  & &
 \hspace*{-20pt}
  \gamma^{}_{\text{vac}}(t)=\lambda^{}_{s}
  \Gamma(s-1)
  \nonumber\\
   & &
 \hspace*{-10pt}
   {}\times
  \left\{
  1- \frac{\cos\left[(s-1)\,\tan^{-1} (\Omega t)\right]}
          {\left(1+\Omega^{2}t^{2}\right)^{(s-1)/2}}\right\},
             \quad (s\not=1),
       \label{Gamma-vac-snot1}\\[5pt]
       & &
 \hspace*{-20pt}
   \gamma^{}_{\text{vac}}(t)=
  \frac{\lambda^{}_{1}}{2}\,\ln\left(1+\Omega^2 t^2\right),
  \quad
  (s=1).
   \label{Gamma-vac-s=1}
   \end{eqnarray}
 \end{subequations}
The latter expression is identical with the well known
result~\cite{BreuerPetr02}.
 Note that the properties of $\gamma^{}_{\text{vac}}(t)$ given by
 Eqs.~(\ref{Gamma-vac-s}) are consistent with the general statements
 formulated in the beginning of this section. First, the vacuum term
 is a monotonically increasing function of time for $s\leq 1$. Second,
  in the superohmic case ($s>1$) this term has a long-time limit:
  \begin{equation}
  \lim_{t\to\infty} \gamma^{}_{\text{vac}}(t)=
   \lambda^{}_{s}\Gamma(s-1),
   \quad
    (s>1).
   \label{Gamma-vac-sup}
  \end{equation}
It is easy to check that this limit is exactly the same as given by
 Eq.~(\ref{Gamma-infty-vac}). One can also see
  from Eq.~(\ref{Gamma-vac-snot1}) that $\gamma^{}_{\text{vac}}(t)$
  monotonically saturates to $\gamma^{}_{\text{vac}}(\infty)$ for
  $1<s\leq 2$ and is a nonmonotonic function of time for
   $s>2$. These  properties of the vacuum term are illustrated in
    Fig.~\ref{Fig-1-r}.

Evaluation of the thermal term (\ref{Gamma(t)-th}) with the spectral
density (\ref{J-model-exp}) also reduces to performing standard integrals.
 After some manipulations which we omit, we obtain
   \begin{widetext}
   \begin{subequations}
      \label{Gamma-th-s}
   \begin{eqnarray}
 \gamma^{}_{\text{th}}(t)&=&
   2\lambda^{}_{s}\left(\Omega\beta\right)^{1-s}\,
   \Gamma(s-1)
     \nonumber\\[5pt]
      & &
    {}\times\sum^{\infty}_{k=1}
   \frac{1}{\left(k+1/\Omega\beta\right)^{s-1}}
   \left\{
   1 -
     \left[1+
    \frac{(t/\beta)^{2}}
         {\left(k+1/\Omega\beta \right)^{2}}\right]^{-(s-1)/2}
    \cos\left[(s-1) \varphi^{}_{k}(t)\right]
   \right\},
   \quad
    (s>0,\ s\not=1),
    \label{Gamma-th-snot1}
    \\[8pt]
      \gamma^{}_{\text{th}}(t)&=& 2\lambda^{}_{1}
  \left[
 \ln \Gamma\left(1+1/\Omega\beta\right)
    -\frac{1}{2}\,\ln\left|\Gamma\left(1+ 1/\Omega\beta
                                       +it/\beta\right)\right|^{2}
  \right],
  \qquad (s=1),
    \label{Gamma-th-s=1}
    \end{eqnarray}
  \end{subequations}
   \end{widetext}
where we have introduced the notation
    \begin{equation}
   \varphi^{}_{k}(t)=
   \tan^{-1}\left(\frac{t/\beta}{k+ 1/\Omega\beta}\right).
    \label{phi-k-snot1}
  \end{equation}

At low temperatures ($\Omega\beta\gg 1 $)
 expression (\ref{Gamma-th-s=1}) reduces to the well known
  result~\cite{BreuerPetr02,Unruh95,Palma96}
     \begin{eqnarray}
    \hspace*{-20pt}
 \gamma^{}_{\text{th}}(t)&=& -\lambda^{}_{1}
   \ln\left|\Gamma(1+it/\beta)\right|^{2}
   \nonumber\\[5pt]
    {}&= & \lambda^{}_{1}\,\ln\left[
  \frac{\sinh(t/\tau^{}_{B})}{t/\tau^{}_{B}}
  \right],
  \quad
   (s=1,\ \Omega\beta\gg 1),
    \label{Gamma-s=1-appr}
  \end{eqnarray}
where
\begin{equation}
  \tau^{}_{B}=\beta/\pi\equiv \hbar/\pi k^{}_{\text{B}} T
   \label{tau-B}
 \end{equation}
   is the so-called thermal correlation time.

We see from Eq.~(\ref{Gamma-th-snot1}) that in the case of incomplete
 decoherence ($s>2$) the thermal term has
 a long-time limit
   \begin{eqnarray}
   & &
   \hspace*{-25pt}
    \lim_{t\to \infty}\gamma^{}_{\text{th}}(t)=
  2\lambda^{}_{s}\left(\Omega\beta\right)^{1-s}
   \nonumber\\
   & &
   \hspace*{5pt}
   {}\times\Gamma(s-1)\, \zeta\!\left(s-1,1+ 1/\Omega\beta\right),
   \quad (s>2),
   \label{Gamma-th-limit-s>2}
  \end{eqnarray}
where $\zeta(z,v)$ is the generalized Riemann zeta function.
 One may easily verify that the above result is
 exactly the same as predicted by Eq.~(\ref{Gamma-infty-th}).
 It is somewhat more difficult to clarify the time behavior of
  $\gamma^{}_{\text{th}}(t)$ directly from Eq.~(\ref{Gamma-th-s}).
  We show in Appendix~\ref{App-C-Der} that, for
   $0<s\leq 2$,  $\gamma^{}_{\text{th}}(t)$
   is a monotonically increasing function of time
    and $\gamma^{}_{\text{th}}(t)\to \infty$
    as $t\to\infty$. In the regime of ``incomplete
    decoherence'' ($s>2$), the thermal term
    exhibits nonmonotonic time behavior for sufficiently large
     values of the parameter $s$
    (see Appendix~\ref{App-C-Der} and Fig.~\ref{Fig-2-r}).

     \begin{figure}[htb] 
\begin{center}
{\includegraphics[width=7.5cm]{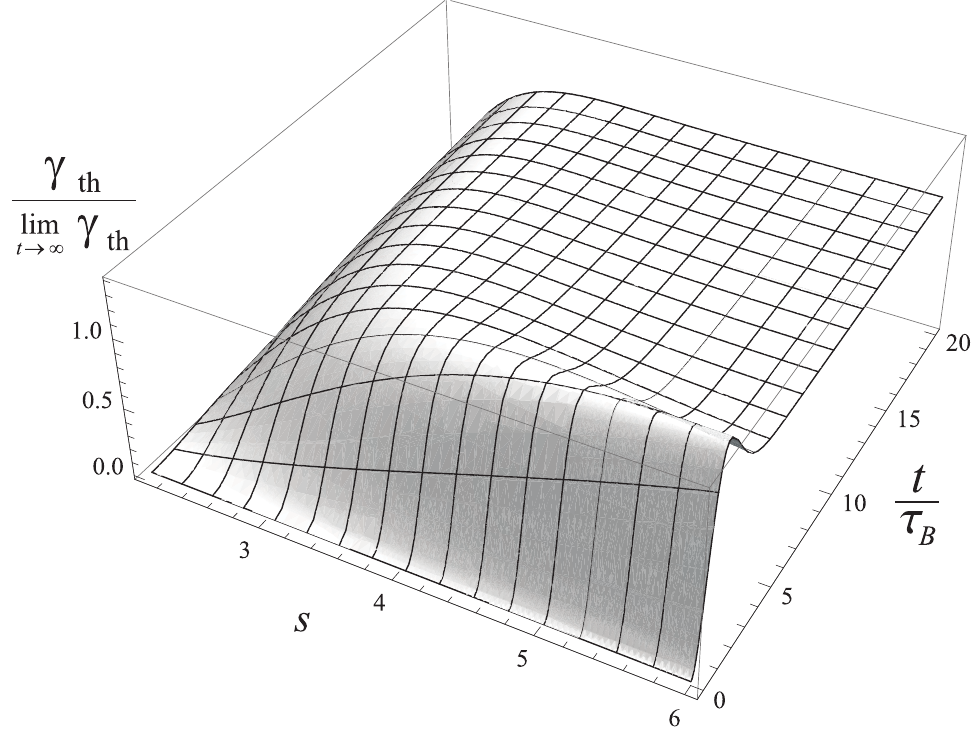}}
 \end{center}
 \caption{Time dependence of the thermal contribution to
  the decoherence function in the regime of ``incomplete decoherence''
   ($s>2$); $\tau^{}_{B}$ is given by Eq.~(\ref{tau-B})
   and $\Omega\tau^{}_{B}=10$.}
 \label{Fig-2-r}
\end{figure}

\begin{figure}[htb] 
\begin{center}
{\includegraphics[width=7.5cm]{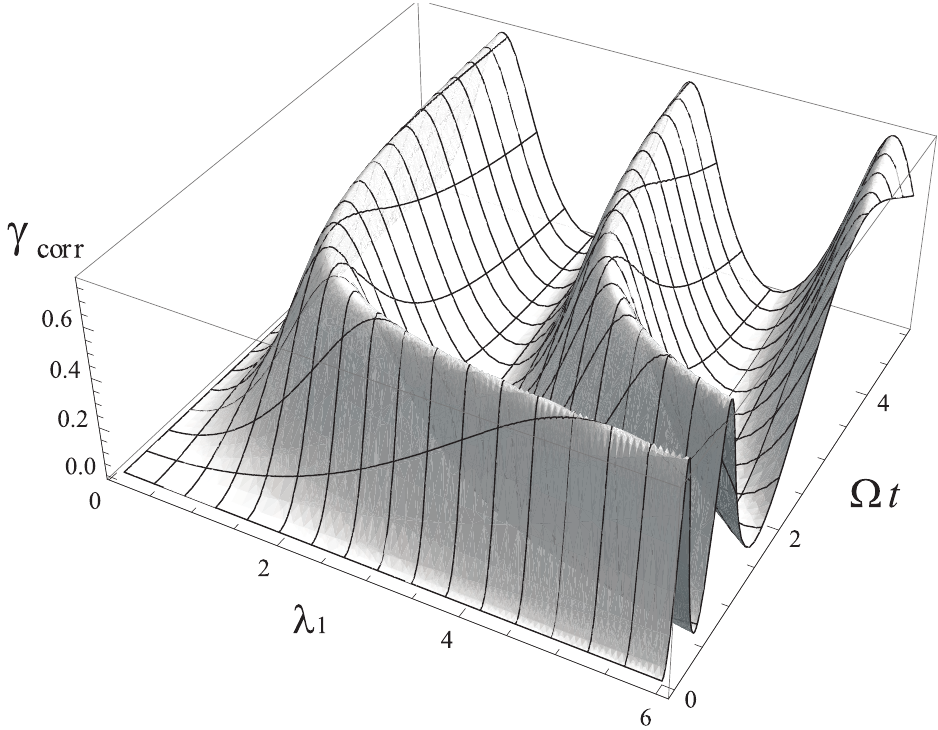}}
 \end{center}
 \caption{Time dependence of the correlation contribution to
  the decoherence function in the ohmic case ($s=1$) for different
   values of the coupling constant $\lambda^{}_{1}$.
  Parameter values: $\langle\sigma^{}_{3}\rangle=0$
  (equal populations
 of the qubit states $|0\rangle$ and $|1\rangle$);
  $\omega^{}_{0}\beta=1$.}
 \label{Fig-3-r}
\end{figure}

We now turn to the correlation term
 (\ref{Gamma-corr}) in the decoherence function assuming the bath spectral
 density to be given by Eq.~(\ref{J-model-exp}). As a preliminary step,
 we calculate the phase function (\ref{Phi-def}). Doing the $\omega$-integral, we
 find
   \begin{subequations}
    \label{Phi-s}
    \begin{eqnarray}
    & &
    \hspace*{-20pt}
  \Phi(t)=
   \frac{\lambda^{}_{s}\Gamma(s-1)}
        {\left(1+ \Omega^{2}t^{2}\right)^{(s-1)/2}}
        \nonumber\\
        & &
        {}\times\sin\left[(s-1)\,\tan^{-1}(\Omega t)\right],
        \quad
        (s>0,\ s\not=1),
   \label{Phi-s-not1}
   \\[5pt]
   & &
    \hspace*{-20pt}
   \Phi(t)=\lambda^{}_{1}\tan^{-1}(\Omega t),
  \quad
  (s=1).
   \label{Phi-s=1}
   \end{eqnarray}
 \end{subequations}
These expressions allow one to bring out some important
 properties of the correlation contribution to the decoherence function.
 First, in the subohmic case ($0<s<1$), the phase function~(\ref{Phi-s-not1})
 increases with time and, consequently,
  $\gamma^{}_{\text{corr}}(t)$ oscillates.
   Second, for $s\geq 1$ the phase function has long-time limits
    \begin{equation}
  \begin{array}{ll}
  \displaystyle
   \lim_{t\to\infty}\Phi(t)= \lambda^{}_{1}\pi/2,
    & \quad (s=1),\\[8pt]
    \displaystyle
 \lim_{t\to\infty}\Phi(t)=0,
    & \quad (s>1),
  \end{array}
  \label{Phi-s-limits}
 \end{equation}
which are consistent with Eq.~(\ref{Phi-infty}). We thus conclude that in
the superohmic case ($s>1$) the correlation term
  (\ref{Gamma-corr}) asymptotically tends to zero as
   $t\to\infty$, whereas
   in the ohmic case ($s=1$) it has a long-time limit
   \begin{eqnarray}
   & &
   \hspace*{-15pt}
  \lim_{t\to \infty}\gamma^{}_{\text{corr}}(t)
   \nonumber\\[5pt]
   & &
   \hspace*{-10pt}
   {}-\frac{1}{2}\,\ln
   \left[
    1 -\frac{\left(1- \langle\sigma^{}_{3}\rangle^{2}\right)
        \sin^{2}(\lambda^{}_{1}\pi/2)
            }
            {\left[\cosh(\beta\omega^{}_{0}/2)-\langle\sigma^{}_{3}\rangle
             \sinh(\beta\omega^{}_{0}/2)\right]^{2}
            }
   \right].
    \label{Gam-corr-lim-s=1}
   \end{eqnarray}
It is interesting to note that this limiting value is a periodic function
of the coupling constant $\lambda^{}_{1}$. In particular, for
 $\lambda^{}_{1}=2n,\ (n=1,2,\ldots)$, we
  have $\gamma^{}_{\text{corr}}(\infty)=0$. The time behavior of
  $\gamma^{}_{\text{corr}}(t)$ in the ohmic case is very sensitive
 to the value of the coupling constant. For sufficiently
   large $\lambda^{}_{1}$,  $\gamma^{}_{\text{corr}}(t)$ has
  a ``peak'' structure (see Fig.~\ref{Fig-3-r}).

Qualitative properties of the contributions to the total decoherence
function $\widetilde{\gamma}(t)$ for different values of the parameter
 $s$ in Eq.~(\ref{J-model-exp})
 are summarized in Table~\ref{Table1}. We note that the value $s=2$ plays a
role of a ``critical parameter'' for decoherence in the model
(\ref{H-tot}). If $s\leq 2$, we have the regime of
 ``complete decoherence'' since
 $\widetilde{\gamma}(t)\to \infty$ as $t\to\infty$, and hence the
  coherences (\ref{sig-tilde}) asymptotically tend to zero.
  For $s>2$, the total decoherence function has a finite
  long-time limit determined by the vacuum and thermal
  contributions. In this case we are dealing  with the
  regime of  ``incomplete decoherence''.
  It should, however,
   be pointed out that in a real qubit
    the residual coherences $\langle\sigma^{}_{\pm}(\infty)\rangle $
 decay to zero due to dissipative processes which are not included
 in the model (\ref{H-tot}).
  \begin{table*}
\caption{\label{Table1} Time behavior of different contributions to the
 total decoherence function $\widetilde{\gamma}(t)$ for the bath spectral
  density (\ref{J-model-exp}).}
\begin{ruledtabular}
\begin{tabular}{cccc}
 {}      & $\gamma^{}_{\text{vac}}(t)$
         &$\gamma^{}_{\text{th}}(t)$
         &$\gamma^{}_{\text{corr}}(t)$ \\ \hline
 $0<s<1$ & Monotonic increase
         & Monotonic increase
         & Oscillations
 \\[5pt]
 $s=1$  & Monotonic increase
        & Monotonic increase
        & ``Peak'' structure;
 \\
        & {}
        & {}
        & $\gamma^{}_{\text{corr}}(\infty)=0$\ or\
         $\gamma^{}_{\text{corr}}(\infty)\not=0$
 \\
        & {}
        & {}
        & depending on the value of $\lambda^{}_{1}$
 \\[5pt]
 $1<s\leq 2$ & Saturation to $\gamma^{}_{\text{vac}}(\infty)\not=0$
             & Monotonic increase
             & Nonmonotonic decay
 \\[5pt]
 $s>2$       & Saturation to $\gamma^{}_{\text{vac}}(\infty)\not=0$
             & Saturation to $\gamma^{}_{\text{th}}(\infty)\not=0$
             & Nonmonotonic decay
 \\
\end{tabular}
\end{ruledtabular}
\end{table*}

 We close this section with remarks about the role
 of initial qubit-bath correlations in different
  regimes of decoherence.
 Some conclusions concerning correlation effects can be
drawn directly from Eq.~(\ref{Gamma-corr}). First,
 in the weak coupling
 limit ($\lambda^{}_{s}\ll 1$)  Eqs.~(\ref{Phi-s}) and (\ref{Gamma-corr}) give
 $\gamma^{}_{\text{corr}}\propto \lambda^{2}_{s}$, while
  $\gamma^{}_{\text{vac}}\propto \lambda^{}_{s}$ and
   $\gamma^{}_{\text{th}}\propto \lambda^{}_{s}$.
  We see that in this limit the main contribution
  to the total decoherence function (\ref{Gamma-tilde})
  is from its ``dynamical'' part  $\gamma(t)$.
  Second, it is clear
   that the correlation term (\ref{Gamma-corr}) is  small
    compared to $\gamma(t)$
   at extremely low temperatures ($\beta\omega^{}_{0}\gg 1$)
 for all values of $\lambda^{}_{s}$. We thus conclude that
  the role of initial qubit-bath correlations becomes
  pronounced in the temperature region
   $\beta\omega^{}_{0} \lesssim 1$ for intermediate strength of coupling.

\begin{figure*}[htb] 
 \begin{center}
{\includegraphics[width=5.6cm]{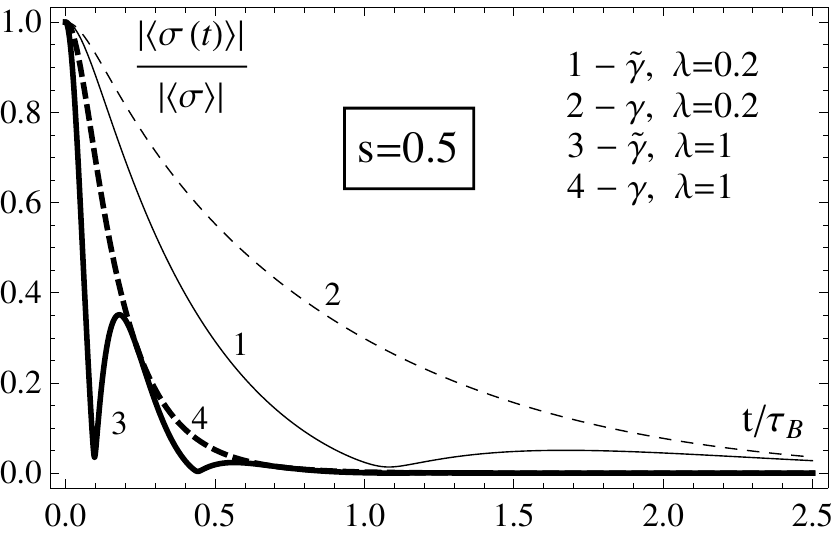}}\hspace*{10pt}
{\includegraphics[width=5.6cm]{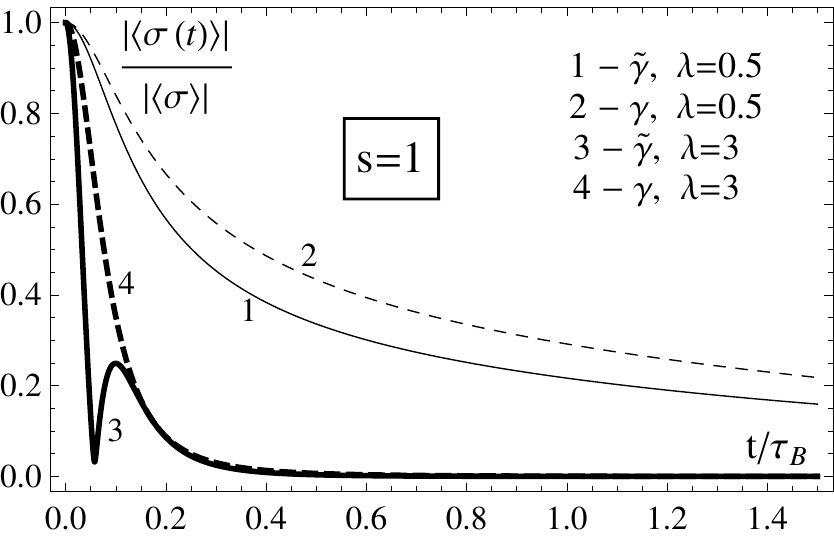}}\hspace*{10pt}
{\includegraphics[width=5.6cm]{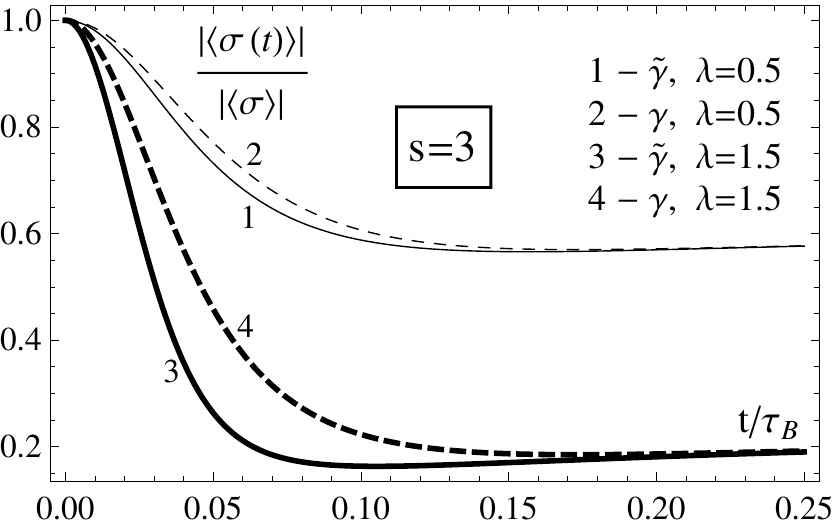}}
\end{center}
 \caption{Time dependence of the coherences
 $\langle\sigma(t)\rangle\equiv\langle\sigma^{}_{\pm}(t)\rangle$
  for  the correlated initial condition
   (\ref{Pure-corr}) (solid lines) and the uncorrelated initial
    condition (\ref{Init-fact}) (dashed lines). In all cases
    $\lambda\equiv \lambda^{}_{s}$;
    $\tau^{}_{B}$ is given by Eq.~(\ref{tau-B}).
  Parameter values: $\Omega\tau^{}_{B}=10$, $\omega^{}_{0}\beta=0.1$.}
 \label{Fig-4-r}
\end{figure*}

 Formulas (\ref{Phi-s}) show that
 $\Omega^{-1}$ determines the characteristic time scale for the correlation
effects. The same is true for the vacuum term~(\ref{Gamma-vac-s}). On the
other hand, the quantity $\tau^{}_{B}$ given by Eq.~(\ref{tau-B})
  determines the time scale for thermal
 effects in decoherence
  [cf.~Eqs.~(\ref{Gamma-th-s}) and (\ref{phi-k-snot1})].
In the most interesting low-temperature range ($\Omega\beta\gg 1$),
 we have $\Omega\tau^{}_{B}\gg 1$, so that
  initial correlations might be expected to  have a pronounced effect on
 the coherences $\langle\sigma^{}_{\pm}(t)\rangle$ at times $t< \tau^{}_{B}$
  when the main contribution to the ``dynamical part'' $\gamma(t)$
   of the decoherence function  comes from vacuum fluctuations in the bath.
For the superohmic case, this can be clearly seen in Fig.~\ref{Fig-4-r}
which illustrates the time behavior of the coherences in different
regimes. Note, however, that in the subohmic and ohmic cases, for
sufficiently weak coupling,
 the correlation effects
 manifest themselves at times $t>\tau^{}_{B}$ where thermal
 excitations dominate (see Fig.~\ref{Fig-4-r}).
 The reason can be traced to the
  facts that,  in the former case,
  the correlation term $\gamma^{}_{\text{corr}}(t)$ exhibits
   undamped oscillations, while in the latter case it has a long-time
   ``plateau''.

\section{Entropy}
Entropy plays a crucial role in the theory of open systems since it is a
natural measure of the lack of information about a system.
 It is thus of interest to discuss the time behavior of the qubit entropy
for the model under consideration.

We start from the
 general expression for the von Neumann (information)
  entropy of a quantum system:
  \begin{equation}
   S(t)= - \text{Tr}^{}_{S}
   \left\{ \varrho^{}_{S}(t)\ln\varrho^{}_{S}(t)\right\}.
   \label{S(t)-def}
  \end{equation}
In the case of a qubit, it is convenient to express
 $S(t)$ in terms of the Bloch vector
  $\vec{v}(t)=\langle\vec{\sigma}(t)\rangle$ using two
  representations
 for the density matrix $\varrho^{}_{S}(t)$. The standard
 representation is~\cite{Allen75,Cohen77}
   \begin{equation}
   \varrho^{}_{S}(t)= \frac{1}{2}\left[
                 1+ \vec{\sigma}\cdot\vec{v}(t)\right].
     \label{rho-v-st}
   \end{equation}
In Appendix~\ref{App-B-Exp} we derive another representation which is
better suited to calculate $\ln\varrho^{}_{S}(t)$:
  \begin{equation}
 \varrho^{}_{S}(t)= \frac{1}{2}\sqrt{1-v^{2}(t)}\,
 \exp\left[
    \vec{\sigma}\cdot \vec{u}(t)
     \right],
    \label{rho-v-esp}
  \end{equation}
where $\vec{u}\upuparrows \vec{v}$, and
  \begin{equation}
  u=\frac{1}{2}\,\ln\left(\frac{1+v}{1-v}\right).
   \label{u-v}
  \end{equation}
Strictly speaking, formula (\ref{rho-v-esp}) is valid only
 if $v<1$, i.e., for a mixed state. Note, however, that the limit
 $v\to 1$ can be taken directly
  in the entropy~(\ref{S(t)-def}) after calculating
 the trace.
 With expressions (\ref{rho-v-esp}) and (\ref{u-v}) one easily derives from
Eq.~(\ref{S(t)-def})
  \begin{equation}
 S(t)= \ln 2
  - \frac{1}{2}\left(1+v\right)\ln\left(1+v\right)
  - \frac{1}{2}\left(1-v\right)\ln\left(1-v\right).
    \label{S-qubit}
  \end{equation}
For a pure state ($v\to 1$) this formula gives $S=0$, as it should be.

The square modulus of the Bloch vector can in general be written as
  \begin{equation}
   v^{2}(t)=
    4\langle \sigma^{}_{+}(t)\rangle \langle \sigma^{}_{-}(t)\rangle
    +\langle \sigma^{}_{3}(t)\rangle^{2}.
    \label{v-sig}
  \end{equation}
For simplicity we shall restrict further discussion to correlated initial
states of the form (\ref{Pure-corr}). Then we have $v(0)=1$ and,
consequently,
 \[
 \langle \sigma^{}_{+}\rangle \langle \sigma^{}_{-}\rangle=
 \frac{1}{4}\left(1- \langle\sigma^{}_{3}\rangle^{2}\right).
 \]
Now using the solution~(\ref{sig-tilde}) and taking into account that
$\sigma^{}_{3}$ is an integral of motion, we obtain
 from Eq.~(\ref{v-sig})
  \begin{equation}
  v(t)=
   \left[ \langle\sigma^{}_{3}\rangle^{2}
   + \left(1 -\langle \sigma^{}_{3}\rangle^{2}\right)
   \text{e}^{-2\widetilde\gamma(t)}\right]^{1/2}.
    \label{v-expl}
  \end{equation}
Formulas (\ref{S-qubit}) and (\ref{v-expl}) determine the time evolution
of the qubit entropy.

In discussing the properties of entropy in the model (\ref{H-tot}) under
the assumption~(\ref{J-model}) for the bath spectral density,
  it is necessary to distinguish two cases:
 the regime of ``complete decoherence'', and the regime of
  ``incomplete decoherence''. In the former case ($s\leq 2$) we have
   $\widetilde{\gamma}(t)\to \infty$ as $t\to\infty$, and hence
    the limiting value of
 the entropy is the same for both (correlated and uncorrelated)
  initial conditions:
   \begin{eqnarray}
   & &
   \hspace*{-30pt}
  \lim_{t\to \infty} S^{}_{}(t)=
   \ln 2
  - \frac{1}{2}\left(1+|\langle\sigma^{}_{3}\rangle|\right)
  \ln\!\left(1+|\langle\sigma^{}_{3}\rangle|\right)
  \nonumber\\
   & &
  {} - \frac{1}{2}\left(1-|\langle \sigma^{}_{3}\rangle|\right)
  \ln\!\left(1-|\langle\sigma^{}_{3}\rangle|\right),
  \quad
  (s\leq 2).
  \label{S-limit-compl}
   \end{eqnarray}
The maximum entropy $S^{}_{\text{max}}(\infty)=\ln 2$ corresponds to
 the initial state with equal populations
 ($\langle\sigma^{}_{3}\rangle=0$). A similar situation
  occurs in the case of ``incomplete decoherence'' ($s>2$) since
   $\gamma^{}_{\text{corr}}(t)\to 0$ as $t\to\infty$. Note, however,
   that the limiting value of the qubit entropy is now given by
   \begin{eqnarray}
   & &
   \hspace*{-20pt}
   \lim_{t\to \infty} S^{}_{}(t)=
   \ln 2 - \frac{1}{2}\left(1+v^{}_{\infty}\right)\ln\left(1+v^{}_{\infty}\right)
   \nonumber\\
  & &
  {}- \frac{1}{2}\left(1-v^{}_{\infty}\right)\ln\left(1-v^{}_{\infty}\right),
  \quad  (s>2),
    \label{S-limit-incompl}
   \end{eqnarray}
where
  \begin{equation}
 v^{}_{\infty}=
   \left[ \langle\sigma^{}_{3}\rangle^{2}
   + \left(1 -\langle \sigma^{}_{3}\rangle^{2}\right)
   \text{e}^{-2\gamma(\infty)}\right]^{1/2}.
   \label{v-infty}
  \end{equation}
Recalling Eqs.~(\ref{Gamma-vac-sup}) and (\ref{Gamma-th-limit-s>2}), we
have
  \begin{eqnarray}
   & &
   \hspace*{-15pt}
  \gamma(\infty)=
  \lambda^{}_{s}\Gamma(s-1)
  \nonumber\\
  & &
  \hspace*{-10pt}
  {}\times\left[
 1+ \frac{2}{\left(\Omega\beta\right)^{s-1}}\,
 \zeta\!\left(s-1,1+ 1/\Omega\beta\right)
  \right],
   \quad (s>2).
    \label{gamma-infty}
  \end{eqnarray}
The second term in braces corresponds to the contribution from thermal
 excitations in the bath and is relatively small
  in the low-temperature range ($\Omega\beta\gg 1$).

 Although initial qubit-bath correlations do not contribute to
 $S(\infty)$ for all $s>0$, they influence the behavior of the
  qubit entropy at times $t<\tau^{}_{B}$. The main reason is easy to see
  when one recalls that, for sufficiently strong coupling,
  the  modulus of the Bloch vector $v(t)$ may have rather sharp
  peaks associated with the evolution of initial correlations
   (see, e.g., Fig.~\ref{Fig-4-r}).

\begin{figure}[h] 
 \begin{center}
\includegraphics[width=7cm]{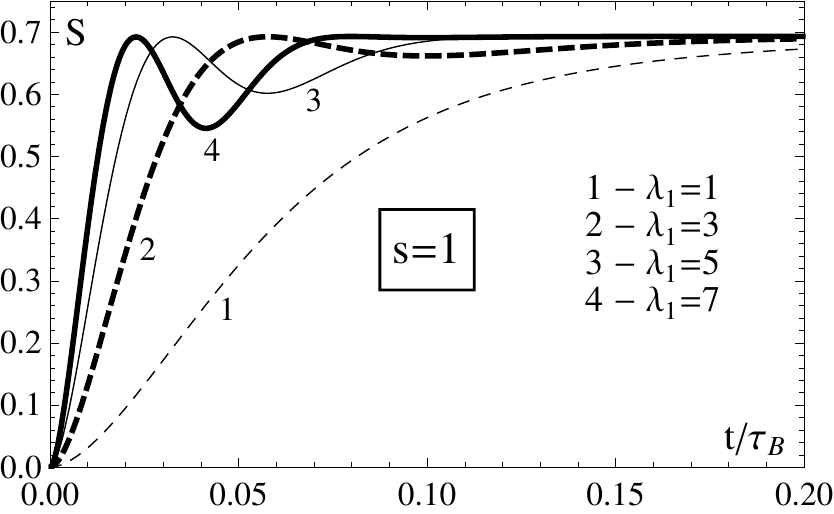}
 \end{center}
 \caption{Time evolution of the qubit entropy in the ohmic case
 for different coupling strengths.
  Parameter values: $\Omega\tau^{}_{B}=10$, $\omega^{}_{0}\beta=0.1$.
  }
 \label{Fig-5-r}
 \end{figure}

    Especially interesting is the
   ohmic case ($s=1$) where the time dependence of the coherences
   is very sensitive to the value of the coupling constant
   $\lambda^{}_{1}$. Figure~\ref{Fig-5-r}
    illustrates the kind of the behavior
    of the qubit entropy one might expect in this case.
 We call attention to the fact that in some time intervals
 the entropy for a larger coupling constant is lower than that for
 a smaller coupling constant.

\section{Conclusions}

Let us present a summary of the results obtained in this work and
 discuss their relation to problems in open system dynamics.

We have derived exact formulas (\ref{sig-fin-gen})
 and (\ref{sig-P}) which describe the time behavior of the coherences
 $\langle\sigma^{}_{\pm}(t)\rangle $
  (off-diagonal elements of the qubit density matrix)
 for all correlated initial states of the form (\ref{Init-corr})
  and (\ref{Pure-corr}),
 respectively. Initial states of these types can be
 interpreted physically in terms of the general theory of quantum
 measurements~\cite{BreuerPetr02,Kraus83,Braginsky92}.
 In particular, operators $\Omega^{}_{m}$ entering
  Eq.~(\ref{sig-fin-gen}) represent
  operations during the ``preparation'' of a combined qubit-bath
  system in an initial state at temperature $T=1/\beta$.

   In this paper we have
 studied in detail the reduced qubit dynamics for the initial condition
 (\ref{Pure-corr}) which corresponds to the
 preparation of the qubit in some pure state $|\psi\rangle$.
  Since the qubit is not isolated from the bath,
  initial correlations come into play through the initial bath density
   matrix (\ref{Bath-psi}) which is a functional of $|\psi\rangle$
   [see also Eq.~(\ref{Bath-psi-expl})].

The quantity of most interest is the decoherence function
$\widetilde{\gamma}(t)$ which determines the decay of the coherences:
$\langle\sigma^{}_{\pm}(t)\rangle
  \propto \exp\left[-\widetilde{\gamma}(t)\right]$.
 Within the framework of the model under consideration, the decoherence
 function can be split into the vacuum part $\gamma^{}_{\text{vac}}(t)$,
 the thermal part $\gamma^{}_{\text{th}}(t)$, and the
  ``correlation'' part $\gamma^{}_{\text{corr}}(t)$.
  In Sec.~IV we have established some connections
   (with different levels of generality) between
    the properties of the bath spectral
     density $J(\omega)$ and possible regimes of decoherence.
First, we have simple but quite general relations (\ref{Gamma-suff})
 and (\ref{Gamma-infty}) which show that the
  long-time behavior of all the contributions to the total decoherence
  function depends  crucially on the form of
  $J(\omega)$ in the low-frequency range. In particular,
   inequalities  (\ref{Gamma-suff}) give sufficient conditions
   for the regime of ``incomplete decoherence'' when
    $\langle\sigma^{}_{\pm}(t)\rangle$ do not tend to zero as
     $t\to\infty$.
      Second, assuming only that in the low-frequency range
  $J(\omega)$ scales as   $J(\omega)\propto \omega^{s}$ with $s>0$, we have shown
   that the regime of ``incomplete decoherence'' occurs for $s>2$ when
   the total decoherence function is bounded.
   Finally,  taking the bath spectral density in the
    form (\ref{J-model-exp}), we have derived exact expressions for
     all the terms in $\widetilde{\gamma}(t)$. As expected, these
     expressions confirm the earlier general predictions.

     We have seen that the qualitative features of decoherence
       are essentially determined by the behavior of $J(\omega)$ in the
      range $\omega\ll \Omega$, where $\Omega$ is a
        cutoff frequency, characteristic of the bath modes. Thus,
      although explicit results have been obtained for a special (exponential)
      form of the cutoff function in $J(\omega)$,
      there is a good reason
      to think that the conclusions summarized in Table~\ref{Table1}
      apply as well to other forms of the cutoff function.

Decoherence is a fundamental property of open quantum systems
  which are met in quantum optics, electronics, atomic and molecular
  physics, etc. The literature concerning diverse aspects of
   this phenomenon is now quite voluminous.
     At first sight a study of very simple models like that specified
     by the Hamiltonian (\ref{H-tot}) might appear to be
     essentially a problem of applied mathematics. In a sense this is
     true. It should be noted, however, that the pure dephasing mechanism
     of decoherence described by the model (\ref{H-tot}) can dominate
     in real physical systems~\cite{Schuster07,Cronin09}. From this point
     of view exactly solvable dephasing models
      seem as themselves to deserve thorough studies. Note also
      that exact results obtained for a dephasing model can serve
      for constructing approximate solutions for more complicated dissipative
      systems~\cite{BreuerPetr02} if the dissipative coupling with
       environment is weak.
 Another field of applications of exactly solvable dephasing
  models with initial system-environment correlations is the theory of
  quantum information.
  As a first step it would be desirable to generalize the results
   to many-qubit systems (quantum registers).

     Recent works~\cite{DajkaLucz10,Laine10} show that initial
      system-environment correlations can play an interesting
      and somewhat unexpected role in decoherence. We hope that
     the results of the present paper will be useful
     for further studies of correlations effects
      in decoherence phenomena.

\begin{acknowledgments}

This work  was supported by  DFG (Deutsche Forschungsgemeinschaft),
 SFB 652 (Sonderforschungsbereich --- Collective Research Center 652).
\end{acknowledgments}

\appendix

\section{Heisenberg equations of motion \label{App-A-Heis}}
Since $\sigma^{}_{3}$ commutes with the Hamiltonian (\ref{H-tot}),
 equations of motion for $b^{}_{k}(t)$ and
  $b^{\dagger}_{k}(t)$ are
   \[
 i\frac{db^{}_{k}(t)}{dt}= \omega^{}_{k}b^{}_{k}(t) +
 \sigma^{}_{3}g^{}_{k},
 \quad
 -i\frac{db^{\dagger}_{k}(t)}{dt}= \omega^{}_{k}b^{\dagger}_{k}(t) +
 \sigma^{}_{3}g^{*}_{k}.
   \]
The solution of these equations with initial conditions
 $b^{}_{k}(0)=b^{}_{k}$ and  $b^{\dagger}_{k}(0)=b^{\dagger}_{k}$
  is easily found to be given by Eqs.~(\ref{b(t)}).
 Equation of motion for  $\sigma^{}_{+}(t)$ reads:
    \[
   i\frac{d\sigma^{}_{+}(t)}{dt}= -\omega^{}_{0}\sigma^{}_{+}(t)
    -2\sum_{k}\left[
            g^{}_{k}b^{\dagger}_{k}(t) + g^{*}_{k}b^{}_{k}(t)
              \right]\sigma^{}_{+}(t).
    \]
Using Eqs.~(\ref{b(t)}) and relation
 $\sigma^{}_{3}\sigma^{}_{+}(t)=\sigma^{}_{+}(t)$,
we get
   \begin{equation}
   i\frac{d\sigma^{}_{+}(t)}{dt}= -\omega^{}_{0}\sigma^{}_{+}(t)
    - W(t)\sigma^{}_{+}(t)
     \label{A-sig-eq}
   \end{equation}
with
  \begin{eqnarray}
   & &
   \hspace*{-20pt}
    W(t)= 2\sum_{k}\Big\{
    g^{}_{k}\text{e}^{i\omega^{}_{k}t}
        \big[
         b^{\dagger}_{k} + \alpha^{*}_{k}(t)/2
        \big]
   \nonumber\\
   & &
   \hspace*{50pt}
   {}+
 g^{*}_{k}\text{e}^{-i\omega^{}_{k}t}
        \left[
         b^{}_{k} + \alpha^{}_{k}(t)/2
        \right]
       \Big\}.
     \label{W(t)}
 \end{eqnarray}
The solution of Eq.~(\ref{A-sig-eq}) is given by
  \begin{equation}
  \sigma^{}_{+}(t)=\exp\left(i\omega^{}_{0}t\right)\,
  \exp^{}_{+}\left[
   i\int^{t}_{0} W(\tau)\,d\tau
      \right]\sigma^{}_{+},
      \label{sig-solut-exp}
  \end{equation}
where $\exp^{}_{+}\left[\ldots\right]$ is the chronologically  ordered
 exponent. Expression (\ref{sig-solut-exp}) can be simplified using the
 operator identity
       \begin{widetext}
   \begin{equation}
  \exp^{}_{+}\left[i\int^{t}_{0} A(\tau)\,d\tau\right]=
  \exp\left\{-
 \frac{1}{2}\int^{t}_{0}d\tau^{}_{1}\int^{\tau^{}_{1}}_{0}
  d\tau^{}_{2}\,
     [A(\tau^{}_{1}),A(\tau^{}_{2})]
     \right\}\,
    \exp\left[i\int^{t}_{0} A(\tau)\,d\tau\right],
     \label{A-identity}
   \end{equation}
   \end{widetext}
which is valid if the commutator $[A(\tau^{}_{1}),A(\tau^{}_{2})]$ at
 different times is a $c$-number function. It is easily verified
 that the operator (\ref{W(t)}) satisfies
   \[
   \left[W(t^{}_{1}),W(t^{}_{2})\right]=
   -8i \sum_{k} |g^{}_{k}|^{2}\,
   \sin[\omega^{}_{k}(t^{}_{1}-t^{}_{2})].
   \]
Applying Eq.~(\ref{A-identity}) to $W$ and evaluating the integrals on the
r.h.s. of that formula, one shows that the $c$-number functions cancel.
Then from Eq.~(\ref{sig-solut-exp})
 follow expressions (\ref{sig-pm(t)}) for $\sigma^{}_{+}(t)$ and
 $\sigma^{}_{-}(t)=\sigma^{\dagger}_{+}(t)$.

 \section{Time derivative of
  $\gamma^{}_{\text{th}}(t)$ \label{App-C-Der}}

Let us consider the time derivative of  $\gamma^{}_{\text{th}}(t)$ given
by Eq.~(\ref{Gamma(t)-th}). We have
        \begin{equation}
 \frac{d\gamma^{}_{\text{th}}}{dt}=
  2\int^{\infty}_{0} d\omega\,
 \frac{J(\omega)}{{\rm e}^{\beta\omega}-1}\,
  \frac{\sin\omega t}{\omega}.
    \label{Gamma-der-th}
   \end{equation}
Under the assumption (\ref{J-model}), this reads
        \begin{equation}
 \frac{d\gamma^{}_{\text{th}}}{dt}=
 \frac{2 \lambda^{}_{s}(\Omega\beta)^{1-s}}{\beta}
 \int^{\infty}_{0} dx\,
 \frac{x^{s-1} F\left(x/\Omega \beta \right)}{{\rm e}^{x}-1}
  \sin(xt/\beta)\, .
    \label{Gamma-der-th-s1}
   \end{equation}
In the limit $\Omega\beta\gg 1$, the cutoff function may be replaced by
 $F(0)=1$. Then, performing the $x$-integral, we get
        \begin{equation}
   \frac{d\gamma^{}_{\text{th}}}{dt}=
   \frac{2 \lambda^{}_{s}(\Omega\beta)^{1-s}}{\beta}\,
   \Gamma(s)
   \sum^{\infty}_{k=1}
   \frac{\sin\left[s\,\tan^{-1}(t/\beta k)\right]}
   {\left[k^{2}+(t/\beta)^{2}\right]^{s/2}},
   \label{Gamma-th-der-low}
  \end{equation}
where $\Gamma(s)$ is the gamma function. If $0<s\leq 2$, we see that
 $d\gamma^{}_{\text{th}}/dt >0$ for any $t>0$.
However, this fact is not sufficient to conclude that the decoherence
  function diverges as $t \to \infty$ since
   it does not exclude the possibility that
    $d\gamma_{\text{th}}/dt \to 0^+$ as $t \to \infty$
    and $\gamma^{}_{\text{th}}(t)$ saturates to a finite value.
 Relation~(\ref{Gamma-infty-th}) excludes  this possibility.
 It tells us that, for $0<s\leq 2$, the decoherence function
  $\gamma^{}_{\text{th}}(t)$ cannot have a
  definite long-time limit since, if it had, the integral
     on the r.h.s. would be finite. We thus conclude that
      $\gamma_{\text{th}}(t)$
     is a monotonically increasing function and tends to
      infinity as $t \to \infty$.

If the bath spectral density is given by Eq.~(\ref{J-model-exp}), the
cutoff function in
 Eq.~(\ref{Gamma-der-th-s1}) has the form
$F(x/\Omega\beta)=\exp\left(-x/\Omega\beta\right)$. Now the integration
over $x$ leads to
   \begin{eqnarray}
    & &
    \hspace*{-25pt}
   \frac{d\gamma^{}_{\text{th}}(t)}{dt}=
   \frac{2 \lambda^{}_{s}(\Omega\beta)^{1-s}}{\beta}\,
   \Gamma(s)
   \\[5pt]
   & &
   \hspace*{-10pt}
   \times{}\sum^{\infty}_{k=1}
   \frac{\sin\left[s\,\varphi^{}_{k}(t)\right]}
   {\left[\left(k+1/\Omega\beta\right)^{2}+ (t/\beta)^{2}\right]^{s/2}},
   \quad
   (s>0),
   \label{Gamma-th-der-Omega}
  \end{eqnarray}
where $\varphi^{}_{k}(t)$ is given by Eq.~(\ref{phi-k-snot1}). Since
 $0<\varphi_{k}(t)<\pi/2$ for all $t>0$, we may employ the same arguments
  as above. Thus, for $0<s\leq 2$, the thermal term
  $\gamma^{}_{\text{th}}(t)$ monotonically increases with time
   and $\gamma^{}_{\text{th}}(t)\to\infty$ as $t\to\infty$.
A more  elaborate analysis of Eqs.~(\ref{Gamma-th-der-low}) and
(\ref{Gamma-th-der-Omega}), which will not be given here, shows that in
the regime of ``incomplete decoherence''
   the derivative $d\gamma^{}_{\text{th}}(t)/dt$ is  a
   positive definite function of time for $2<s\leq 3$ and may change sign
   for $s>3$. In the latter case the thermal term
    $\gamma^{}_{\text{th}}(t)$ exhibits nonmonotonic time behavior.

\section{Exponential form of the density matrix \label{App-B-Exp}}

The derivation of formula (\ref{rho-v-esp}) is based on the following
 property of the Pauli matrices (see Problem 1 to \S~55 in
 Ref.~\cite{LandLif77}).
  Let $f$ be some function of the operator
  $a+\vec{\sigma}\cdot\vec{u}$, where $a$ and $\vec{u}$ are
  $c$-number quantities. Then
   \begin{equation}
   f(a+\vec{\sigma}\cdot\vec{u})= A + \vec{\sigma}\cdot\vec{B},
 \label{f-lin}
   \end{equation}
where
 \begin{equation}
  \begin{array}{l}
  \displaystyle
  A=\frac{1}{2}\left[
  f(a+u) + f(a-u)\right],
  \\[7pt]
   \displaystyle
  \vec{B}= \frac{1}{2}\,\frac{\vec{u}}{u}
   \left[
   f(a+u) -f(a-u)
  \right],
   \end{array}
  \label{AB}
  \end{equation}
and $u=|\vec{u}|$. Identifying  the r.h.s. of Eq.~(\ref{f-lin}) with
 the qubit density matrix written in the form (\ref{rho-v-st}),  we
 have $A=1/2$ and $\vec{B}=\vec{v}/2$.
  On the l.h.s. of Eq.~(\ref{f-lin}) we take
 $f(x)=\exp(x)$. Then Eqs.~(\ref{AB}) give
  \begin{eqnarray*}
   1=\text{e}^{a}\left(\text{e}^{u}+ \text{e}^{-u}\right),
   \qquad
   v=\text{e}^{a}\left(\text{e}^{u}- \text{e}^{-u}\right).
  \end{eqnarray*}
Solving these equations for $a$ and $u$, we find
  \begin{eqnarray*}
    \text{e}^{a}= \frac{1}{2}\,\sqrt{1-v^2},
   \qquad
   u= \frac{1}{2}\,\ln\left(\frac{1+v}{1-v}\right).
  \end{eqnarray*}
This leads immediately to
 the representation (\ref{rho-v-esp}) for the density matrix.


\begin{thebibliography}{00}
 \bibitem{BreuerPetr02}
 H.-P. Breuer and F. Petruccione,
 \textit{The Theory of Open Quantum Systems\/}
 (Oxford University Press, Oxford, 2002).
 \bibitem{Weiss08}
 U. Weiss,
 \textit{Quantum Dissipative Systems\/}
 (World Scientific, Singapore, 1999).
 \bibitem{Nakajima58}
 S. Nakajima, Prog. Theor. Phys. \textbf{20}, 948 (1958).
 \bibitem{Zwanzig60}
 R.~Zwanzig, J. Chem. Phys. \textbf{33}, 1338 (1960).
 \bibitem{VaccBreuer10}
 B.~Vacchini and H.-P. Breuer, Phys. Rev. A \textbf{81}, 042103 (2010).
 \bibitem{SmirneVacc10}
 A.~Smirne and B.~Vacchini, Phys. Rev. A \textbf{82}, 022110 (2010).
 \bibitem{Steane98}
 A.~Steane, Rep. Prog. Phys. \textbf{61}, 117 (1998).
 \bibitem{Bouwmeester00}
  D.~Bouwmeester, A.~Ekert, and A.~Zeilinger (eds.),
   \textit{The Physics of Quantum Information\/}
   (Springer-Verlag, Berlin, 2000).
   \bibitem{Valiev05}
  K.A.~Valiev, Phys.-Usp. \textbf{48}, 1 (2005).
  \bibitem{DajkaLucz10}
 J.~Dajka and J.~{\L}uczka, Phys. Rev. A \textbf{82}, 012341 (2010).
  \bibitem{Luczka90}
  J.~{\L}uczka, Physica A \textbf{167}, 919 (1990).
   \bibitem{Unruh95}
  W.G.~Unruh, Phys. Rev. A \textbf{51}, 992 (1995).
   \bibitem{Palma96}
  G. M.~Palma, K.-A.~Suominen, and A.K.~Ekert,
  Proc. R. Soc. London, Ser.~A \textbf{452}, 567 (1996).
   \bibitem{Doetsch74}
  G.~Doetsch,
  \textit{ Introduction to the Theory and Application
 of the Laplace Transformation\/}
  (Springer, Berlin, 1974).
   \bibitem{Legett87}
  A.J.~Leggett \textit{et al.\/}, Rev. Mod. Phys. \textbf{59}, 1 (1987).
   \bibitem{Allen75}
  L.~Allen and J.H.~Eberly,
  \textit{Optical Resonance and Two-level Atoms\/}
 (Dover, N.Y., 1975).
  \bibitem{Cohen77}
  C.~Cohen-Tannoudji, B.~Diu, and F.~Lalo\"e,
  \textit{Quantum Mechanics\/}  (Wiley, N.Y., 1977).
  \bibitem{LandLif77}
  L.D.~Landau and E.M.~Lifshitz,
  \textit{Quantum Mechanics\/} (Pergamon Press, Oxford, 1977).
  \bibitem{Kraus83}
  K.~Kraus, \textit{States, Effects, and Operations\/}
  Vol. 190 of \textit{Lecture Notes in Physics\/}
   (Springer, Berlin, 1983).
   \bibitem{Braginsky92}
  V.B.~Braginsky and F.Ya.~Khalili,
  \textit{Quantum Measurement\/}
  (Cambridge University Press, Cambridge, 1992).
  \bibitem{Schuster07}
  D.I.~Schuster \textit{et al.}, Nature (London) \textbf{445}, 515 (2007).
  \bibitem{Cronin09}
  A.D.~Cronin, J.~Schmiedmayer, and D.E.~Pritchard,
  Rev. Mod. Phys. \textbf{81}, 1051 (2009).
  \bibitem{Laine10}
  E.-M.~Laine, J.~Piilo, and H.-P.~Breuer,
  EPL \textbf{92}, 60010 (2010).
\end{thebibliography}
\end{document}